# Climate-Adaptive and Cascade-Constrained Machine Learning Prediction for Sea Surface Height under Greenhouse Warming

**Tianmu Zheng[1], Ru Chen[1]*, Xin Su[1], Julian Mak[2], Gang Huang[3,4] and Bingzheng Yan[1]**

[1]Tianjin Key Laboratory for Marine Environmental Research and Service, School of Marine Science and Technology, Tianjin University, Tianjin, China.

[2]Department of Ocean Science, The Hong Kong University of Science and Technology, Hong Kong, China.

[3]National Key Laboratory of Earth System Numerical Modeling and Application, Institute of Atmospheric Physics, Chinese Academy of Sciences, Beijing, China.

[4]University of Chinese Academy of Sciences, Beijing, China.

Corresponding author: Ru Chen (ruchen@tju.edu.cn)

**Key Points:**

- A Dual-Attention ConvLSTM model, with a kinetic energy cascade constraint, is developed for sea surface height prediction.
- The machine learning model generalizes well from the present-day climate to greenhouse-warming conditions.
- Incorporating the kinetic energy cascade constraint improves machine learning performance in both current and greenhouse warming climates.

## Abstract

Machine learning (ML) has achieved remarkable success in climate and marine science. Given that greenhouse warming fundamentally reshapes ocean conditions such as stratification, circulation patterns and eddy activity, evaluating the climate adaptability of the ML models is crucial. While physical constraints have been shown to enhance the performance of ML models, kinetic energy (KE) cascade has not been used as a constraint despite its importance in regulating multi-scale ocean motions. Here we develop two sea surface height (SSH) prediction models (with and without KE cascade constraint) and quantify their climate adaptability at the Kuroshio Extension. Both models exhibit only slight performance degradation under greenhouse warming conditions. Incorporating the KE cascade as a physical constraint significantly improves the model performance, reducing eddy kinetic energy errors by 14.7% in the present climate and 15.9% under greenhouse warming. Additional validations using satellite observations and in the Gulf Stream region further confirm the robustness of the proposed models. Compared with the KE spectrum constraint, both constraints improve the cross-scale transfer and spectrum of KE, but the KE cascade constraint yields larger improvements in the cross-scale transfer. This work presents the first application of the KE cascade as a physical constraint for ML-based ocean state prediction and demonstrates its robust adaptability across climates, offering guidance for the further development of global ML models for both present and future conditions.

## Plain Language Summary

Predicting sea surface height (SSH) is crucial for understanding ocean behavior, especially in the context of global warming and rising sea levels. In this study, we developed two machine learning (ML) models for predicting SSH in both current and future climate conditions. By incorporating a novel physical constraint, the kinetic energy (KE) cascade, into the models, we significantly enhanced prediction accuracy. The KE cascade describes how energy transfers across different spatial scales in the ocean, and integrating this concept improves the model's ability to capture the complex behavior of ocean currents and eddies. Our results show that the models remain robust under greenhouse warming, with the KE cascade constraint further improving their predictive performance. We further tested the models using satellite altimetry data and in the Gulf Stream, and compared the cascade-constrained model with the spectrum-constrained model. This work introduces a new ML approach for ocean state prediction and demonstrates how ML can complement traditional climate models in both state forecasts and dynamical analysis.

# 1 Introduction

Machine learning (ML) has demonstrated superior prediction accuracy and computational efficiency in Earth system modeling (Dramsch, 2020; Eyring et al., 2024; Karpatne et al., 2018; Reichstein et al., 2019; Rolnick et al., 2022). For example, models like Pangu-Weather and FuXi have achieved remarkable prediction skill and efficiency in weather forecasting (Bi et al., 2023; Chen et al., 2023). However, it is essential to assess whether these ML models retain reasonable prediction skill under greenhouse warming conditions (Beucler et al., 2024), a capability hereafter referred to as climate adaptability. Furthermore, incorporating physical constraint has been shown to enhance ML model performance (Kashinath et al., 2021; Zhu et al., 2022); however, the kinetic energy (KE) cascade has not yet been explored as a constraint. In this study, we develop a dual-attention ConvLSTM model, quantify the degree of its climate adaptability, and also assess the potential of incorporating the KE cascade to improve the prediction skill.

Climate adaptability is critical for ML-based ocean prediction, as greenhouse warming fundamentally alters ocean dynamics, including stratification, circulation patterns and eddy activity (Fox-Kemper, 2021; Li et al., 2020; Sallée et al., 2021; S. Wang et al., 2024). These changes affect both large-scale circulation and mesoscale eddies (Beech et al., 2022; Martínez-Moreno et al., 2021; Peng et al., 2022). Under greenhouse warming simulated by AWI-CM-1-1-MR under the SSP3-7.0 scenario (Beech et al., 2022; Eyring et al., 2016; O'Neill et al., 2016; Semmler et al., 2020), the geostrophic velocity at the Kuroshio Extension decreases by 11.5%, the mean kinetic energy (MKE) drops by 46.0%, whereas eddy kinetic energy (EKE) rises by 33.1% (Fig. 1). However, existing ocean ML models are typically trained and evaluated on temporally continuous datasets within the same climate state (Bi et al., 2023; Chen et al., 2023; Chen et al., 2024; Ham et al., 2019; Ling et al., 2022; Rasp et al., 2018; Zhang et al., 2023). For example, Xihe was trained on GLORYS12 from 1993-2019 then evaluated over 2019-2020 (X. Wang et al., 2024). Although the community has started to explore the climate adaptability of ML models for atmospheric studies (Hernanz et al., 2022; Kochkov et al., 2024; O'Gorman & Dwyer, 2018) and oceanic parameterization problems (Guillaumin & Zanna, 2021; Gultekin et al., 2024; Perezhogin et al., 2025), this issue has received little attention in ocean prediction research. Therefore, our first aim is to quantify the performance degradation of ML models trained under the current climate when applied to greenhouse warming scenarios.

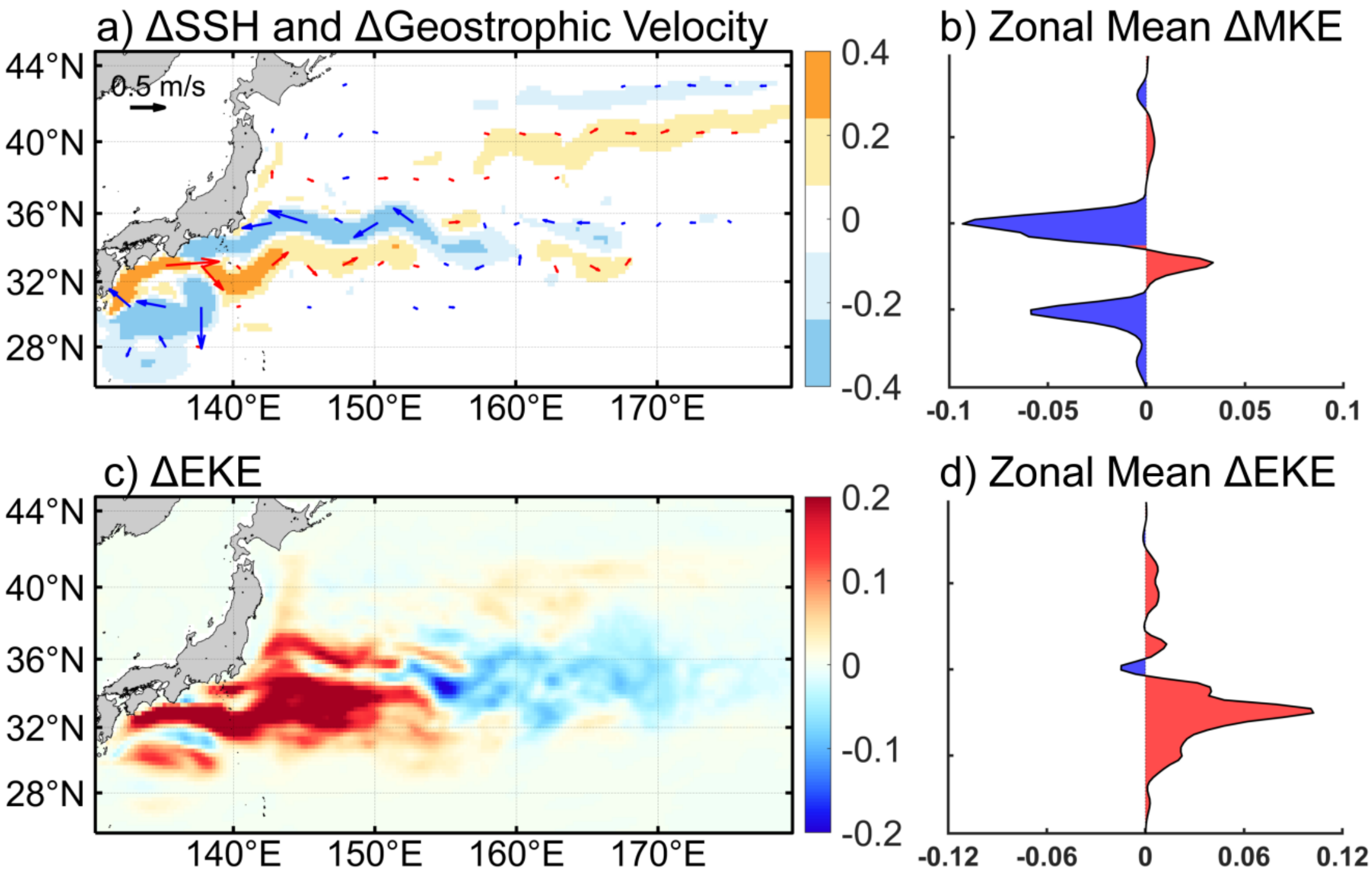

**Figure 1.** Oceanic responses to greenhouse warming. Δ represents the difference between greenhouse warming and current climate from the AWI-CM-1-1-MR model under the SSP370 scenario. (a)Sea surface height (SSH) changes (m, shading) and geostrophic velocity changes (arrows). (b-d) Changes in KE: (b) zonal mean MKE, (c) EKE spatial pattern, (d) zonal mean EKE.

Our second goal is to investigate whether incorporating KE cascade as a physical constraint can enhance the ML model performance. Integrating physical constraint significantly improves the ML model performance in varying conditions (Kashinath et al., 2021; Yuval & O'Gorman, 2020; Zhu et al., 2022). Inspired by these studies, we hypothesize that incorporating appropriate physical constraint could enhance the ML models performance across different climate states.

Here we propose to use geostrophic KE cascade as a physical constraint. The ocean contains multi-scale motions, such as ocean circulation, jets and eddies (Fig. 2). KE is redistributed across a range of spatial scales and is transferred among them through nonlinear interactions—a process known as the KE cascade (Alexakis & Biferale, 2018; Kolmogorov, 1991; Vallis, 2017). KE cascade plays a vital role in modulating multi-scale energy transfers and contributing to climate variability (Sérazin et al., 2018; Storer et al., 2023). However, to the best of our knowledge, it has not been used as a physical constraint of ML models.

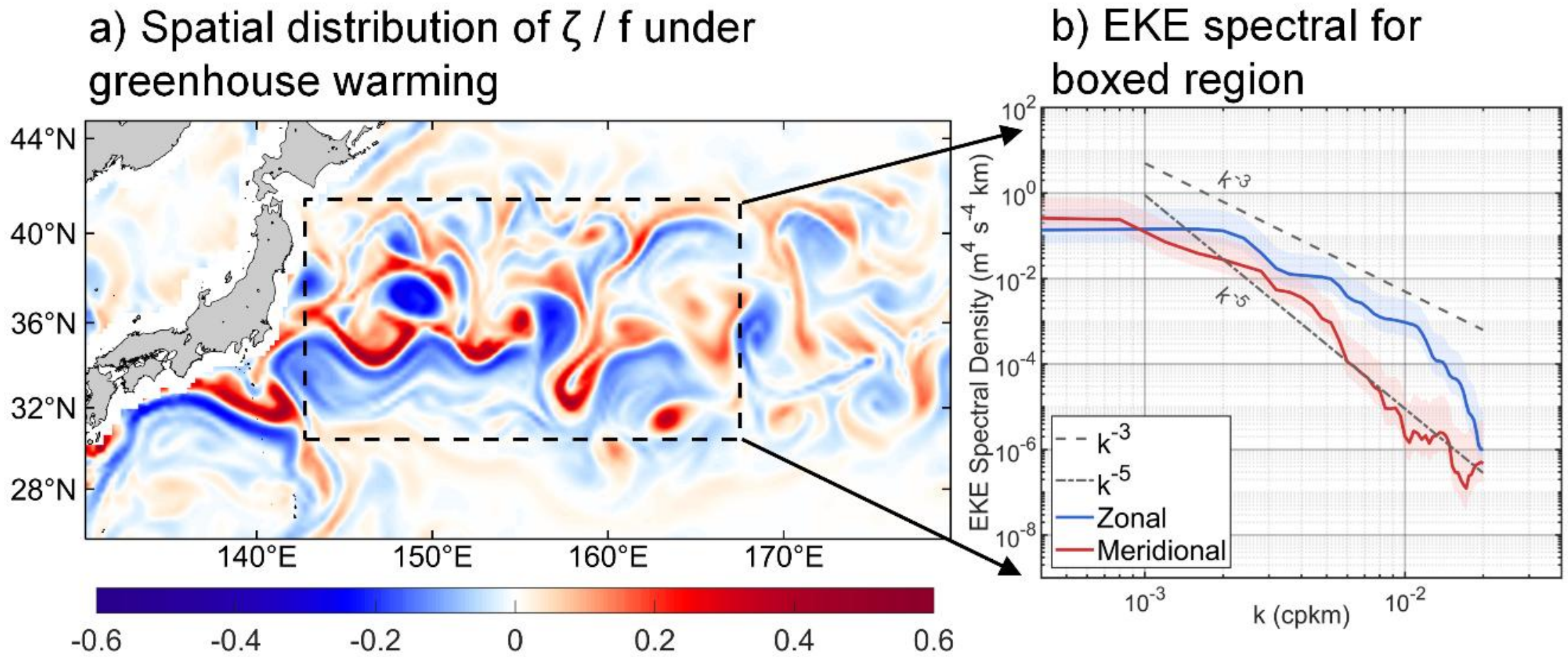

**Figure 2.** Multi-scale ocean motions and energy distribution at the Kuroshio Extension. (a) Spatial distribution of normalized relative vorticity (ζ/f) under greenhouse warming. (b) Wavenumber spectra of EKE for the boxed region. Blue and red lines represent zonal and meridional spectra with 95% confidence intervals (shaded). Gray dashed lines show $k^{-3}$ and $k^{-5}$ reference slopes.

To summarize, our goals are twofold. First, we evaluate the extent to which the ML models maintain adaptability across present and future climate conditions. Second, we investigate whether KE cascade constraint can enhance model performance. We focus on sea surface height (SSH) prediction at the Kuroshio Extension. SSH is a key oceanic variable containing multi-scale dynamical signals, including large-scale circulation and meso- to submesoscale eddies (Chelton et al., 2011; Stammer, 1997). The Kuroshio Extension is an ideal region for this study due to its pronounced SSH spatiotemporal variability and strong sensitivity to climate change (Su et al., 2018; Zhou & Cheng, 2022). Our results show that the ML models we developed for SSH prediction remain robust under greenhouse warming, and that integrating KE cascade constraint further improves their prediction skill. Rostness test using alternative datasets and regions has also been conducted. We also compare the KE cascade constraint with the KE spectrum constraint to assess their respective impacts on both the KE flux and the KE spectrum.

## 2 Machine Learning Method

We develop two models to demonstrate its climate adaptability and the advantage of incorporating KE cascade constraint: 1) the Dual-Attention-ConvLSTM Model (DAM, Section 2.1), and 2) the Cascade-Constrained-ConvLSTM Model (CCM, Section 2.2). The experimental setup is described in Section 3.2.

## 2.1 Architecture of DAM Model

DAM integrates both spatial and temporal attention mechanisms within the ConvLSTM framework to enhance SSH prediction capabilities. While ConvLSTM has proven effective in various spatiotemporal prediction tasks, including precipitation nowcasting, ocean wave forecasting, and wind speed prediction (Shi et al., 2015; Song et al., 2022; Zheng et al., 2023), and attention mechanisms have demonstrated remarkable capability in capturing important features and dependencies in sequential data (Bahdanau et al., 2014; Lin et al., 2020; Vaswani et al., 2017), the ML algorithm integrating both methods have not been designed and used in ocean prediction problems.

The spatial attention module employs average and max pooling with convolutional operations to focus on regions with strong SSH variability. The temporal attention mechanism identifies critical time steps for prediction through learned weighting, enabling the model to capture both spatial and temporal structures crucial for ocean prediction.

Besides the dual-attention mechanisms, DAM incorporates a novel loss function that captures SSH evolution patterns beyond traditional numerical accuracy metrics (Figure 3b). $\mathcal{L}_{DAM}$ is formulated as:

$$\mathcal{L}_{DAM} = \alpha\mathcal{L}_{MSE} + \beta\mathcal{L}_{trend} + \gamma\mathcal{L}_{magnitude} \quad (1)$$

where:

$$\mathcal{L}_{MSE} = \mathrm{MSE}(SSH^{pred}, SSH^{real}) \quad (2)$$

$$\mathcal{L}_{trend} = \mathrm{MSE}(SSH^{pred}_{diff}, SSH^{real}_{diff}) \quad (3)$$

$$\mathcal{L}_{magnitude} = \mathrm{MSE}(SSH^{pred}_{diff}, SSH^{real}_{diff}) \times \mathrm{weight} \quad (4)$$

MSE represent mean squared error, $SSH^{*}_{diff} = SSH^{*}_{t+1} - SSH^{*}_{t}$ represent temporal differences between consecutive time steps. $*^{pred}$ represents the predicted data from the model and $*^{real}$ represents the true data. The weight term assigns higher importance to large variations:

$$\mathrm{weight} = 1 + 2\mathbb{I}(|SSH^{\mathrm{real}}_{\mathrm{diff}}| > 0.1) \quad (5)$$

where $\mathbb{I}$ is the indicator function that equals 1 when the condition is true and 0 otherwise. This weighting mechanism ensures that regions experiencing significant changes (>0.1m) receive enhanced attention during training. We set $\alpha = 0.5$, $\beta = 0.25$, and $\gamma = 0.25$, which were optimized based on validation set performance, to balance numerical accuracy with temporal evolution patterns and magnitude sensitivity.

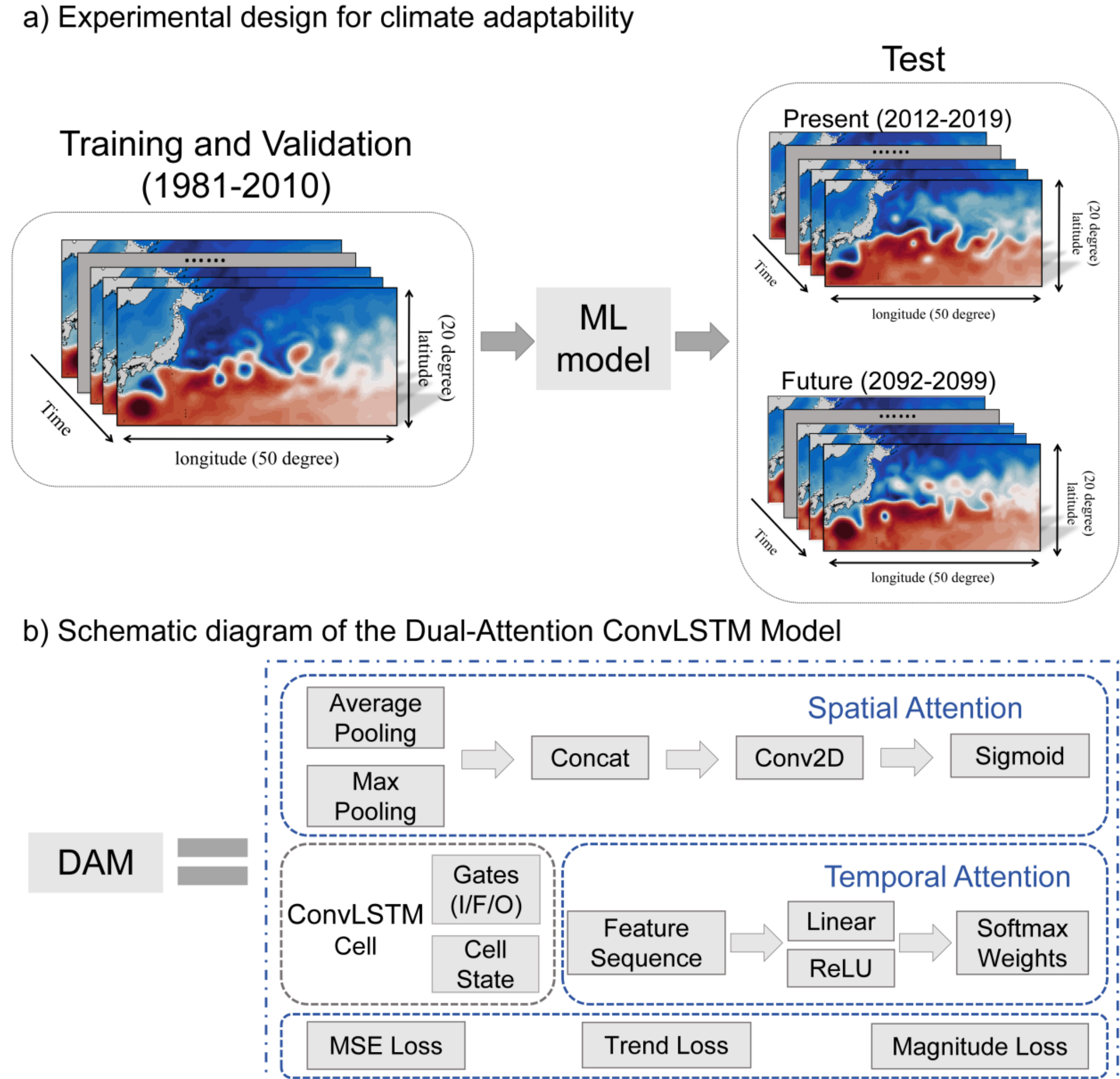

**Figure 3.** Climate adaptability evaluation framework and DAM model architecture. (a) Experimental design for evaluating climate adaptability. The model is trained on the 1981-2010 data and tested on both present (2012-2019) and greenhouse warming (2092-2099) climate conditions. (b) Architecture of the Dual-Attention-ConvLSTM Model (DAM) model, integrating spatial and temporal attention mechanisms within ConvLSTM cells and employing a Trend-Aware Loss function combining MSE, trend, and magnitude components.

### 2.2 Architecture of CCM Model

While physics-informed approaches have incorporated various oceanic constraints (Meng et al., 2023; Meng et al., 2021; Zhu et al., 2022), the KE cascade representing cross-scale energy transfers has not been utilized. Here we incorporate the spectral KE flux as a constraint into our model.

Building upon DAM, we develop CCM that incorporates KE cascade into the neural network architecture (Figure 4). The cascade-constraint loss measures the spectral KE flux error:

$$\mathcal{L}_{cascade} = MSE(\Pi^{pred}, \Pi^{true}) \tag{6}$$

where the spectral KE fluxes ($\Pi$) using the coarse-graining approach (Aluie et al., 2018):

$$\Pi^{g}(x,y;t;\ell) = -\rho_0 \left[ \tau_\ell(u^g,u^g)\frac{\partial u_\ell^g}{\partial x} + \tau_\ell(u^g,v^g)\left(\frac{\partial u_\ell^g}{\partial y} + \frac{\partial v_\ell^g}{\partial x}\right) + \tau_\ell(v^g,v^g)\frac{\partial v_\ell^g}{\partial y} \right] \tag{7}$$

where $\rho_0$ = 1027.4 kg/m$^3$, $\tau_\ell(u,u)$ represents the Reynolds stress tensor, $u^g$ and $v^g$ represent the zonal and meridional geostrophic velocity, and $u_\ell^g$ , $v_\ell^g$ represent the geostrophic velocity components at scales larger than $\ell$ . The Reynolds stress tensor is defined as:

$$\tau_\ell(\mathbf{u},\mathbf{u}) = (\mathbf{uu})_\ell - \mathbf{u}_\ell \mathbf{u}_\ell \tag{8}$$

where this general form is applied to the geostrophic velocity components in equation (7). This tensor quantifies the force exerted by small-scale motions (scales $< \ell$) on large-scale motions (scales $> \ell$). We evaluate $\Pi^g$ at five representative scales (5°, 3°, 2°, 1°, 0.5°) by coarse-graining approach (Aluie et al., 2018). The first three loss components (MSE, trend, and magnitude losses) are identical to those defined in the DAM model. The total CCM loss becomes:

$$\mathcal{L}_{CCM} = \alpha\mathcal{L}_{MSE} + \beta\mathcal{L}_{trend} + \gamma\mathcal{L}_{magnitude} + \delta\mathcal{L}_{cascade} \tag{9}$$

We set $\alpha = 0.4$, $\beta = 0.15$, $\gamma = 0.15$ and $\delta = 0.3$, which were optimized based on validation set performance, to balance numerical accuracy with physical consistency, ensuring the model captures not only "appearance" to true data values but also maintains "intrinsic" physical KE cascade processes. In addition, we develop a Spectrum-Constrained model (SCM) that shares the same architecture as CCM but replaces the cascade loss with a KE spectrum loss, allowing us to evaluate the effects of constraining energy distribution rather than energy transfer.

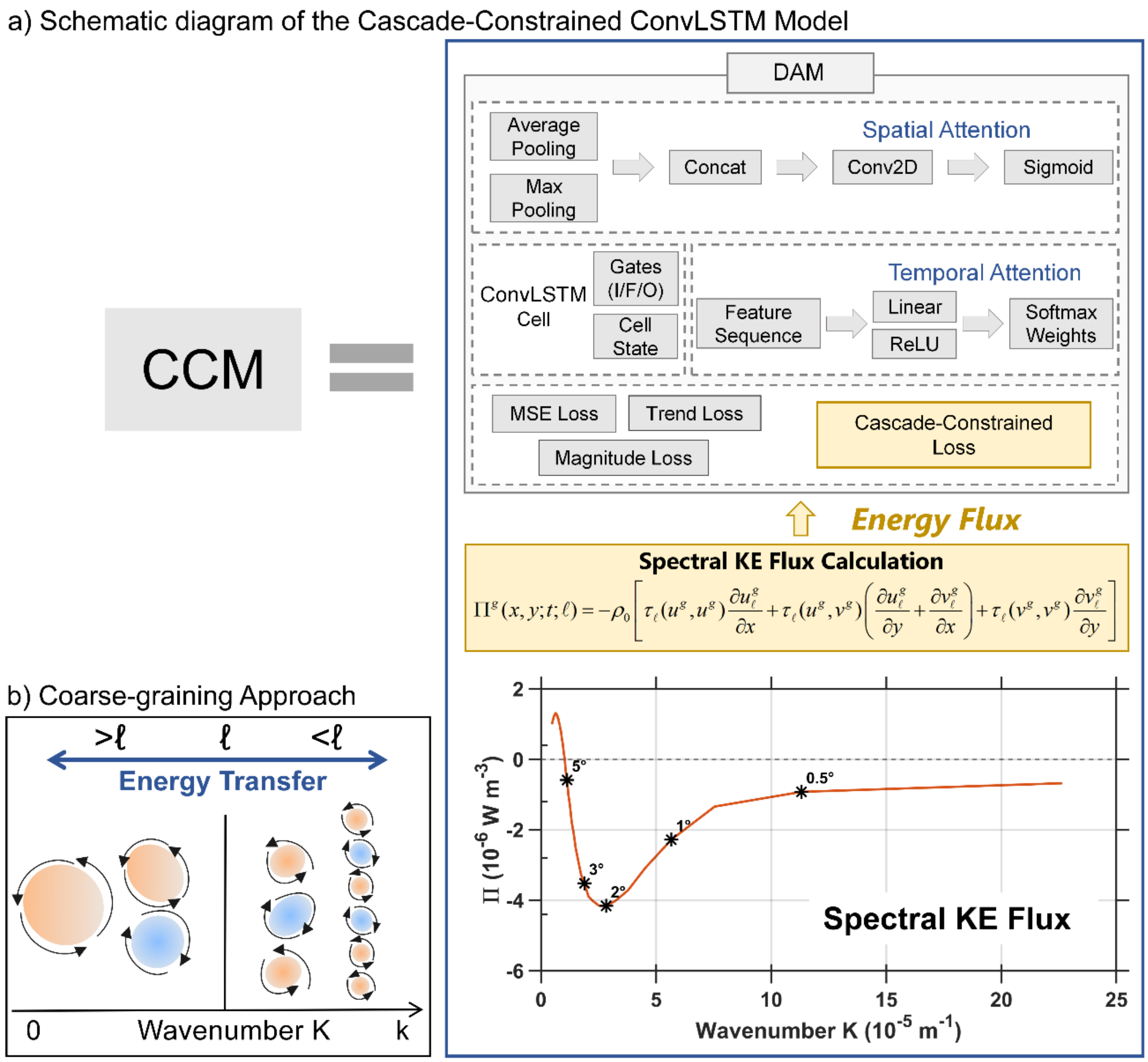

**Figure 4.** Architecture of the Cascade-Constrained-ConvLSTM (CCM) model. (a) Schematic diagram showing the integration of DAM components (spatial attention, temporal attention, and trend-aware loss) with KE cascade constraint. The spectral KE flux Π(k) displays the inverse cascade with our selected filter scales (5°, 3°, 2°, 1°, 0.5°) marked by stars, which are used for computing the physics-informed loss. (b) The spectral KE flux denotes the KE transfer rate between larger-scale (larger than the separation scale ℓ) and smaller-scale motions. Here we apply the coarse-graining approach to estimate the spectral KE flux (Aluie et al., 2018).

# 3 Experimental Framework

## 3.1 Description of Data

The model output analyzed in this study is from the AWI-CM-1-1-MR coupled climate model (Semmler et al., 2020; Semmler et al., 2019), which contributes to the Coupled Model Intercomparison Project Phase 6 (CMIP6) intercomparison project. The ocean component of AWI-CM-1-1-MR is based on the Finite Element Sea Ice-Ocean Model (FESOM) version 1.4 (Danilov et al., 2004), configured with an unstructured triangular mesh that allows for variable horizontal resolution (Wang et al., 2014). The mesh resolution ranges from approximately 8km in dynamically active regions to about 80km (Semmler et al., 2020), with enhanced resolution in areas of high eddy activity (Sein et al., 2017).

The daily SSH fields analyzed here are from the r1i1p1f1 ensemble member both historical runs (1850-2014) and future projections following the SSP3-7.0 scenario (2015-2100). The SSP3-7.0 scenario is a high-emission pathway under which atmospheric $CO_2$ concentrations are projected to approximately double by the end of the century (Eyring et al., 2016; Masson-Delmotte et al., 2021; O'Neill et al., 2016).

The performance of AWI-CM-1-1-MR in simulating ocean dynamics and mesoscale eddy activity has been systematically evaluated (Beech et al., 2022). Comparison with satellite altimetry data (AVISO) shows that AWI-CM-1-1-MR captures approximately 82% of the observed EKE at the Kuroshio Extension, and the model successfully simulates the anticipated intensification of EKE in the Kuroshio Extension under greenhouse warming (Beech et al., 2022). Given its capability to capture cross-scale energy transfers from large scales to mesoscale eddies, the AWI-CM-1-1-MR dataset is particularly suitable for the development and evaluation of our cascade-constrained ML algorithm.

## 3.2 Experiment Setup

We designed four experiments (Figure 3a and Table 1) to investigate climate adaptability and the advantage of using the KE cascade constraint. Through these experiments, we can evaluate the climate adaptability of the ML models we develop and assess whether incorporating the KE cascade constraint helps improve prediction accuracy and retain climate adaptability.

DAM-C vs DAM-F comparison evaluates the climate adaptability of the basic ML model by quantifying performance degradation when a model trained on current climate is applied to greenhouse warming conditions (Section 4.1). The comparison of DAM-C vs CCM-C and DAM-F vs CCM-F demonstrates the advantage of incorporating KE cascade constraint in reducing prediction errors (Section 4.2).

**Table 1.** Experimental setup in this study.

| Experiment Name | Setup | Model | Description |
| --- | --- | --- | --- |
| DAM-C | Train:1981-2010, Test:2012-2019 | DAM | DAM trained on current climate and tested on current climate |
| DAM-F | Train:1981-2010, Test:2092-2099 | DAM | DAM trained on current climate and tested on future climate |
| CCM-C | Train:1981-2010, Test:2012-2019 | CCM | CCM trained on current climate and tested on current climate |
| CCM-F | Train:1981-2010, Test:2092-2099 | CCM | CCM trained on current climate and tested on future climate |

**Note:** All experiments use 21 days SSH input to predict 7 days SSH output.

### 3.3 Evaluation Metrics

this study evaluates the prediction skill of the ML models using both conventional and advanced metrics. Conventional metrics include the normalized root-mean-square error (NRMSE) of SSH, while physical and higher-order statistical metrics—such as information entropy, geostrophic EKE, skewness, and kurtosis—capture aspects like extreme SSH events, the intensity of geostrophic eddy activity, and SSH uncertainty (Table 2. In addition, the geostrophic KE spectral flux is employed as a complementary metric, with its calculation formula shown in Figure 8. Unlike prior studies that focused mainly on conventional metrics, our expanded suite of metrics provides a more complete assessment of ML model performance.

**Table 2.** Metrics for SSH analysis and evaluation.

| Metric | Physical Means |
|---|---|
| $\text{Information entropy}(i,j) = -\sum_{k=1}^{K} p_k(i,j)\log_2 p_k(i,j)$ | Quantifies SSH uncertainty by measuring probability distribution uniformity; higher values indicate greater uncertainty and variability (Shannon, 1948). |
| $\mathrm{EKE}_g(i,j) = \frac{1}{2}[(u_g'(i,j))^2 + (v_g'(i,j))^2]$ | Quantifies the intensity of geostrophic eddy motions (Chelton et al., 2011; Stammer, 1997). |
| $\text{Skewness}(i,j) = \frac{\overline{(\mathrm{SSH}'(i,j))^3}}{\sigma^3(i,j)}$ | Asymmetry of SSH distribution: positive values indicate right-tail, and vice versa (Sura & Gille, 2010; Von Storch & Zwiers, 2002). |
| $\text{Kurtosis}(i,j) = \frac{\overline{(\mathrm{SSH}'(i,j))^4}}{\sigma^4(i,j)} - 3$ | Frequency of extreme SSH events: high values indicate more frequent extremes, and vice versa (Sura & Gille, 2010; Von Storch & Zwiers, 2002). |
| Variable Definitions | |
| • $(i,j)$: Represent spatial coordinate indices<br>• $N$ : Total number of data points<br>• $K$ : Total number of bins (K=10, equally-spaced bins spanning SSH range [-0.3, 1.7] m)<br>• k: Bin index (k = 1, 2, ..., K)<br>• $p_k$ : Probability of SSH values in the k-th bin<br>• $\overline{(\cdot)}$ : Time-averaged<br>• $\mathrm{SSH}'$ : SSH deviation from the time average<br>• $\sigma$ : Standard deviation of $\mathrm{SSH}'$ | • $u_g$ , $v_g$ : Eastward and northward geostrophic velocities:<br>$u_g = -\frac{g}{f}\frac{\partial \mathrm{SSH}}{\partial y}$<br>$v_g = \frac{g}{f}\frac{\partial \mathrm{SSH}}{\partial x}$<br>Where $g$ is gravitational acceleration, $f$ is the Coriolis parameter, $x$ and $y$ are the longitudinal and latitudinal coordinates |

NRMSE is used to evaluate the SSH itself [Eq. (10)], while the physical and statistical metrics are assessed using the Mean Absolute Error (MAE) [Eq. (11)] and correlation R [Eq. (12)]. The NRMSE, MAE and correlation R are defined as follows:

$$\mathrm{NRMSE}_{SSH} = \frac{\sqrt{\frac{1}{N}\sum_{(i,j)}(SSH^{\mathrm{pred}}(i,j) - SSH^{\mathrm{real}}(i,j))^2}}{\max SSH^{\mathrm{real}}(i,j) - \min SSH^{\mathrm{real}}(i,j)} \qquad (10)$$

$$MAE = \frac{1}{N}\sum_{(i,j)}|X^{real}(i,j) - X^{pred}(i,j) \quad (11)$$

$$R = \frac{\sum_{(i,j)}(X^{real}(i,j) - \overline{X^{real}})(X^{pred}(i,j) - \overline{X^{pred}})}{\sqrt{\sum_{(i,j)}(X^{real}(i,j) - \overline{X^{real}})^2 \sum_{(i,j)}(X^{pred}(i,j) - \overline{X^{pred}})^2}} \quad (12)$$

where $X$ represents the physical and statistical metrics listed in Table 2, $(i, j)$ represent spatial coordinate indices, and N denotes the total number of grid points, $*^{pred}$ represents the predicted data from the model and $*^{real}$ represents the true data.

## 4 Result: Climate Adaptability and Advantage of Cascade Constraints

### 4.1 Climate Adaptability of DAM Model

The Kuroshio Extension undergoes intense changes under greenhouse warming, making it an idealized region for climate adaptability evaluation. In the context of greenhouse warming, the EKE band intensifies and shifts westward, getting broader and less coherent. Similarly, the high information entropy region expands both meridionally and zonally, reflecting the broadening of eddy-active zones into previously quiescent regions (Figs. 5a-d). The change of skewness and kurtosis reflects shifts in the distribution asymmetry and extreme event statistics of SSH (Figs. 4a-d). The spatial patterns of these metrics under current climate are consistent with previous studies (Beech et al., 2022; Sura & Gille, 2010; Von Storch & Zwiers, 2002), the EKE response to greenhouse warming also aligns with trends in literature (Barceló-Llull et al., 2025; Beech et al., 2022).

We evaluated the climate adaptability of the DAM model by comparing current climate case (DAM-C) and future climate case (DAM-F) experiments (Table1). Across all the evaluated metrics, the Probability Density Functions (PDFs) of biases remain tightly clustered around zero, demonstrating reasonable performance under altered oceanic conditions (Figs. 5e-f, 6e-f). Compared with the DAM-C, the kurtosis bias in DAM-F exhibits an even narrower PDFs centered near zero, indicating improved rather than degraded performance under greenhouse warming (Fig. 6f). For the skewness (information entropy) bias, the PDFs show a slight shift toward positive (negative) values, although their overall structures remain largely unchanged (Figs. 5f, 6e). These results indicate that the DAM model maintains robust performance across different climate states.

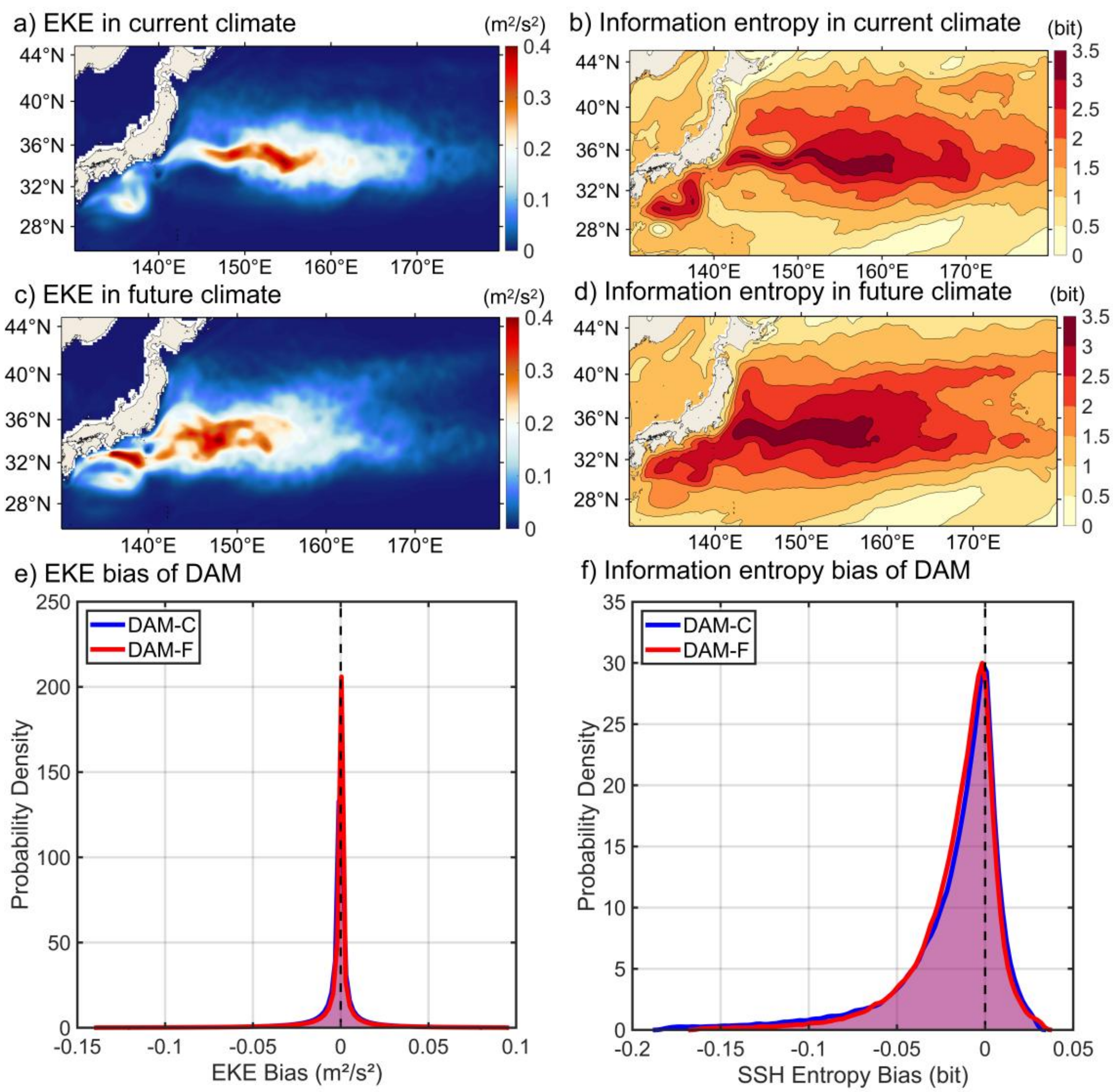

**Figure 5.** Spatial patterns and biases of the time-mean EKE and information entropy for the DAM model. (a) EKE ($m^2/s^2$) in current climate (2012-2019). (b) Information entropy of SSH (bit) in current climate. (c) EKE in future climate (2092-2099) under greenhouse warming. (d) Information entropy of SSH in future climate. (e) PDFs of the EKE bias for current climate (blue) and future climate (red). (f) PDFs of the information entropy bias for current and future climate. The dashed vertical lines in (e) and (f) indicate the line of zero bias.

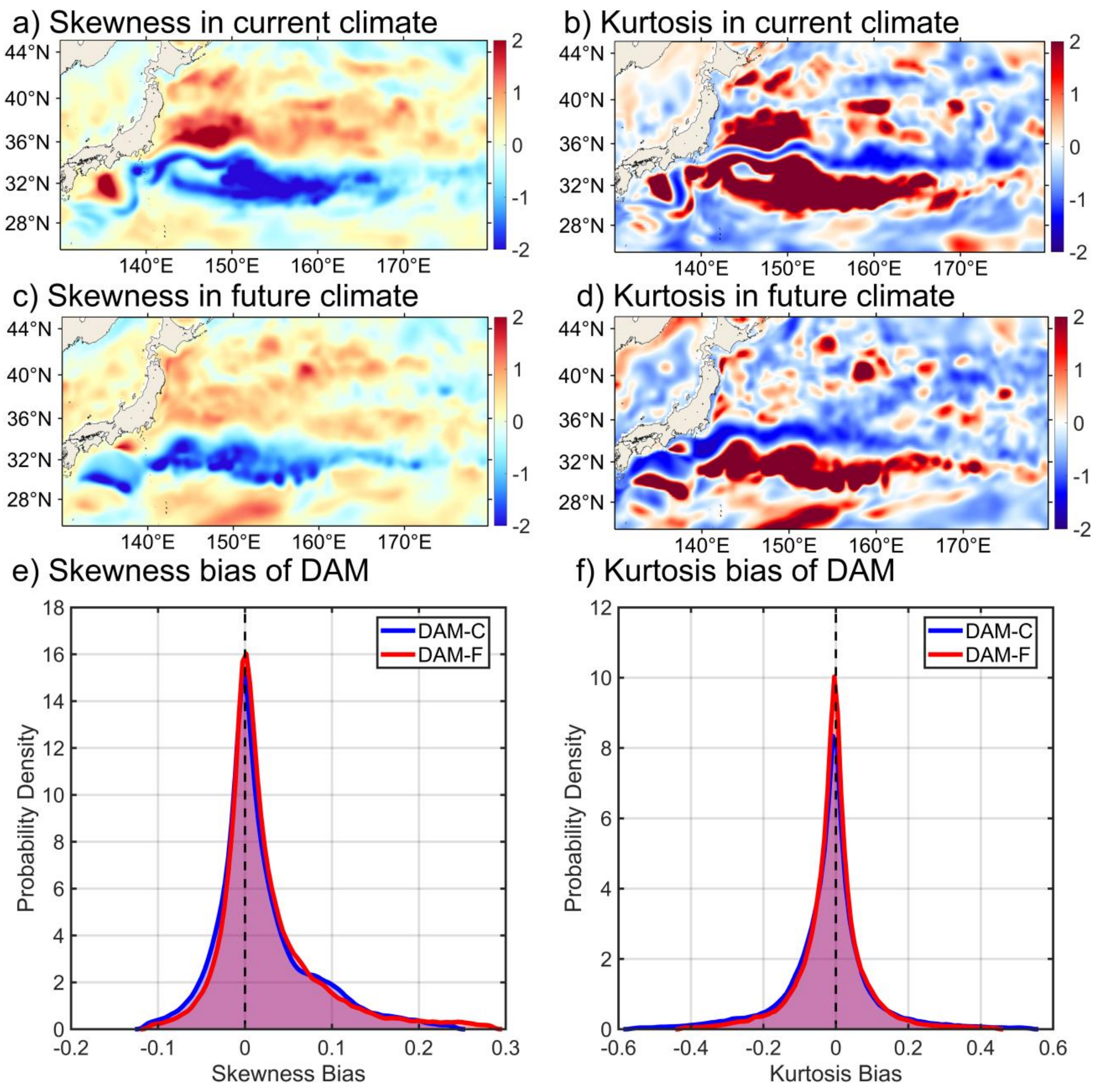

**Figure 6.** Spatial patterns and biases of the time-mean skewness and kurtosis for the DAM model. (a) Skewness and (b) kurtosis in current climate (2012-2019). (c) Skewness and (d) kurtosis in future climate (2092-2099) under greenhouse warming. (e) PDFs of the skewness bias for current (blue) and future climate (red). (f) PDFs of the kurtosis bias for current and future climate conditions. The dashed vertical lines in panels (e) and (f) indicate the line of zero bias.

To quantitatively evaluate the climate adaptability of the DAM model, we compared the metrics (Table 2) described between the DAM-C and DAM-F experiments. The relative changes in the MAE for information entropy and kurtosis are negative, indicating improved prediction skill under greenhouse warming. Similarly, DAM-F (future-climate experiment) shows slightly better skill in representing the spatial structure of information entropy and EKE. Although other metrics in Table 3 exhibit modest degradation in the greenhouse warming case, the errors remain within acceptable bounds. For example, SSH NRMSE increases moderately from 0.0124 to 0.0133, corresponding to only a 7.3% decline in performance. EKE exhibits the largest relative degradation at 18.9%, and skewness shows degradation of only 2.3%. Notably, the spatial pattern correlations between true and predicted values remain consistently high (>0.93) across all metrics. Overall, these results demonstrate that the DAM model maintains robust performance despite altered oceanic conditions, indicating a high degree of climate adaptability.

**Table 3.** Performance comparison between the DAM-C (current climate) and DAM-F (future climate) experiments.

| | DAM-C | DAM-F | Relative change |
|---|---|---|---|
| Normalized Root-Mean-Square Error [Eq. (10)] | | | |
| SSH (m) | 0.0124 | 0.0133 | **+7.3%** |
| Mean Absolute Error [Eq. (11)] | | | |
| Information Entropy (bit) | 0.025 | 0.022 | **-12.0%** |
| EKE ($m^2/s^2$) | 0.0095 | 0.0113 | **+18.9%** |
| Skewness | 0.043 | 0.044 | **+2.30%** |
| Kurtosis | 0.097 | 0.077 | **-20.6%** |
| Spatial Correlation Between True and Predicted Values [Eq. (12)] | | | |
| Information Entropy | 0.9987 | 0.9988 | **+0.01%** |
| EKE | 0.932 | 0.939 | **+0.75%** |
| Skewness | 0.996 | 0.994 | **-0.20%** |
| Kurtosis | 0.993 | 0.989 | **-0.40%** |

**Note:** + performance degradation; **-** performance improvement.
Relative change = (DAM-F - DAM-C) / DAM-C × 100%.

### 4.2 Advantage of Incorporating KE Cascade Constraint

To evaluate the impact of incorporating KE cascade constraint, we compared experiments with and without the constraint (see Table 1 for experiment descriptions). The key message is that adding this constraint consistently reduces the prediction errors across all the evaluation metrics.

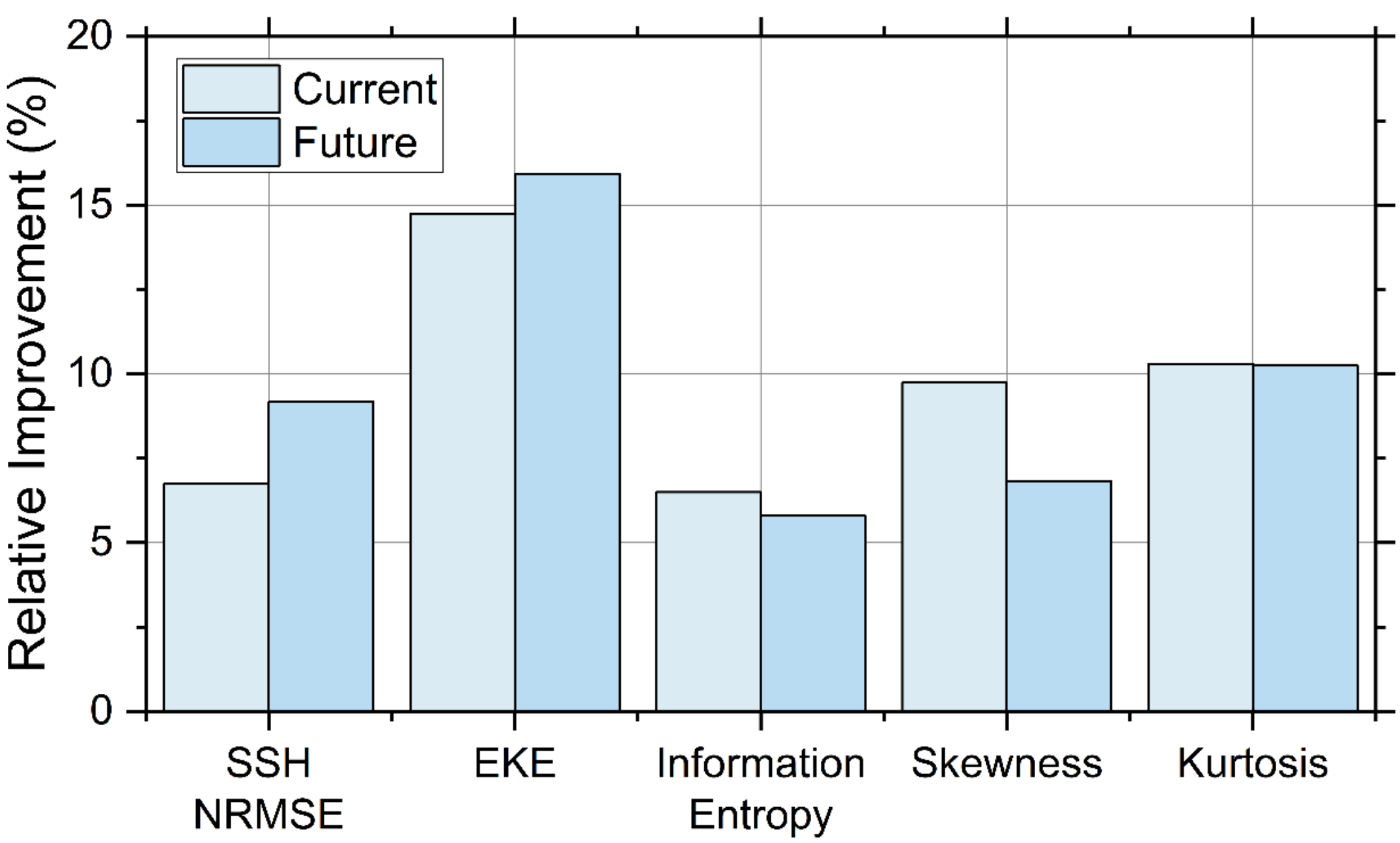

**Figure 7.** Performance comparison between DAM and CCM models showing the time-mean relative improvement by incorporating KE cascade constraint. Bar charts display the relative improvement (%) of CCM over DAM for five evaluation metrics: NRMSE of SSH, EKE MAE, information entropy MAE, skewness MAE, and kurtosis MAE. Light blue bars represent current climate conditions (2012-2019), while dark blue bars represent future climate conditions (2092-2099). The relative improvement is calculated as: (DAM-C - CCM-C) / DAM-C $\times$ 100% for current climate and (DAM-F - CCM-F) / DAM-F $\times$ 100% for future climate. All the improvements are positive, indicating consistent enhancement of prediction skill when KE cascade constraint are incorporated across both climate states.

For both current and future climate scenarios, incorporating KE cascade constraint (CCM) notably improves prediction skill compared to experiments without these constraint (DAM). As shown in Figure 7, under the current climate, CCM-C exhibits 6.76% lower SSH NRMSE values than DAM-C (0.01159 vs 0.01243). Under greenhouse warming, this advantage becomes even more pronounced, with CCM-F exhibiting 9.18% lower errors than DAM-F (0.01207 vs 0.01329). Incorporating this cascade constraint boosts the EKE prediction skill by 14.74% under the current climate and 15.93% under greenhouse warming. Kurtosis also shows substantial improvements with over 10% error reduction in both climate conditions. Information entropy and skewness demonstrate consistent improvements ranging from 5.80% to 9.74%.

The KE cascade depicts cross-scale transfer among multi-scale motions (Alexakis & Biferale, 2018; Kolmogorov, 1991; Vallis, 2017). The performance of this cascade constraint, especially in improving EKE, can be attributed to their ability to enforce physically consistent energy transfers, which greatly modulate mesoscale eddy energy (Geng et al., 2025). By constraining cross-scale KE fluxes during training, CCM also better captures the KE cascade characteristics.

**Table 4.** Spatial correlations between true and predicted metrics in each experiment.

| | | DAM-C | CCM-C | DAM-F | CCM-F |
|---|---|---|---|---|---|
| Metric | Information Entropy | 0.9987 | **0.9989** | 0.9988 | **0.9989** |
| | EKE | 0.932 | **0.951** | 0.939 | **0.957** |
| | Skewness | 0.996 | **0.997** | 0.994 | **0.995** |
| | Kurtosis | 0.993 | **0.994** | 0.989 | **0.991** |

The spatial correlations between true and predicted values further confirm the superior performance of the cascade-constrained algorithm (CCM, Table 4). Although correlation coefficients are high (>0.93) across all metrics and experiments (Table 4), the CCM experiments consistently enhance spatial pattern accuracy beyond the already strong performance of the DAM model. EKE exhibits the most substantial improvements, with correlations rising from 0.932 to 0.951 under the current climate and from 0.939 to 0.957 under greenhouse warming. The remaining metrics also exhibit consistent enhancements (Table 4), highlighting the importance of preserving physically consistent energy transfers in ML ocean models.

## 5 Discussion

### 5.1 Validation in AVISO and Gulf Stream regions

To further assess the robustness of the DAM and CCM models, we evaluate their performance against AVISO satellite altimetry in the Kuroshio Extension. The models presented in Sections 4.1 and 4.2 are applied directly to AVISO without additional retraining. Both models retain high skill in the AVISO evaluation, and CCM shows improvements over DAM. Therefore, analyses based on both CMIP6 and AVISO show that adding the cascade constraint improves the prediction skill of SSH and related dynamical metrics, particularly reducing errors in geostrophic EKE.

This improvement by adding cascade constraints also holds in other energetic regions. For example, we examined regional transferability of the DAM and CCM models by conducting experiments in the Gulf Stream, another energetic western boundary current. Because the land–sea mask differs from that in the Kuroshio Extension, we retrain both DAM and CCM for the Gulf

Stream configuration using the same training strategy described above. Both models achieve high skill in the Gulf Stream evaluation, and CCM generally improves over DAM across most metrics (Table 5).

**Table 5.** Model performance of DAM and CCM across AVISO and Gulf Stream validations.

| | Performance of DAM | | | Performance of CCM | | |
|---|---|---|---|---|---|---|
| | DAM-C | DAM-C (AVISO) | DAM-C (Gulf) | CCM-C | CCM-C (AVISO) | CCM-C (Gulf) |
| Normalized Root-Mean-Square Error [Eq. (10)] | | | | | | |
| SSH (m) | 0.0124 | 0.0148 | 0.0121 | 0.0116 | 0.0132 | 0.0119 |
| Mean Absolute Error [Eq. (11)] | | | | | | |
| Information Entropy (bit) | 0.025 | 0.025 | 0.027 | 0.023 | 0.023 | 0.026 |
| EKE ($m^2/s^2$) | 0.0095 | 0.0119 | 0.0061 | 0.0081 | 0.0104 | 0.0059 |
| Skewness | 0.043 | 0.047 | 0.034 | 0.039 | 0.040 | 0.032 |
| Kurtosis | 0.097 | 0.100 | 0.099 | 0.087 | 0.089 | 0.097 |
| Spatial Correlation Between True and Predicted Values [Eq. (12)] | | | | | | |
| Information Entropy | 0.9987 | 0.9959 | 0. 9958 | 0.9989 | 0.9956 | 0. 9962 |
| EKE | 0.932 | 0.903 | 0.943 | 0.951 | 0.923 | 0.946 |
| Skewness | 0.996 | 0.995 | 0.996 | 0.997 | 0.996 | 0.996 |
| Kurtosis | 0.993 | 0.985 | 0.992 | 0.994 | 0.987 | 0.994 |

**Note:** DAM-C and CCM-C are defined in Table 1. DAM-C (AVISO) denotes validation on AVISO altimetry without retraining, i.e., the DAM-C models trained on the AWI-CM-1-1-MR dataset are directly evaluated on AVISO; CCM-C (AVISO) is defined in the same way. DAM-C (Gulf) denotes evaluation in the Gulf Stream region using models retrained under the DAM-C experiment for the Gulf Stream configuration; CCM-C (Gulf) is defined in the same way.
For the information-entropy metric in the Gulf Stream experiments, the SSH binning range is adjusted to [-1.3, 0.7] m to better match the SSH range in the Gulf Stream.

### 5.2 KE Cascade versus Spectrum Constraints

The spectral KE flux $\Pi(k)$ depicts the KE transfer rate across selected spatial scales (Figs. 8a, b). The negative value of $\Pi(k)$ indicates inverse cascade, where energy flows from small to large scales. The spectral flux diagnosed here (Figs. 8a, 8b) closely resembles the AVISO satellite observations (Geng et al., 2025; Scott & Wang, 2005; Wang et al., 2015). Comparison between Figs. 8a and 8b suggests that the inverse cascade amplitude intensifies under greenhouse warming, consistent with findings in literature (Geng et al., 2025). These results suggest that the AWI-CM-1-1-MR model realistically captures cross-scale energy transfer at the Kuroshio Extension.

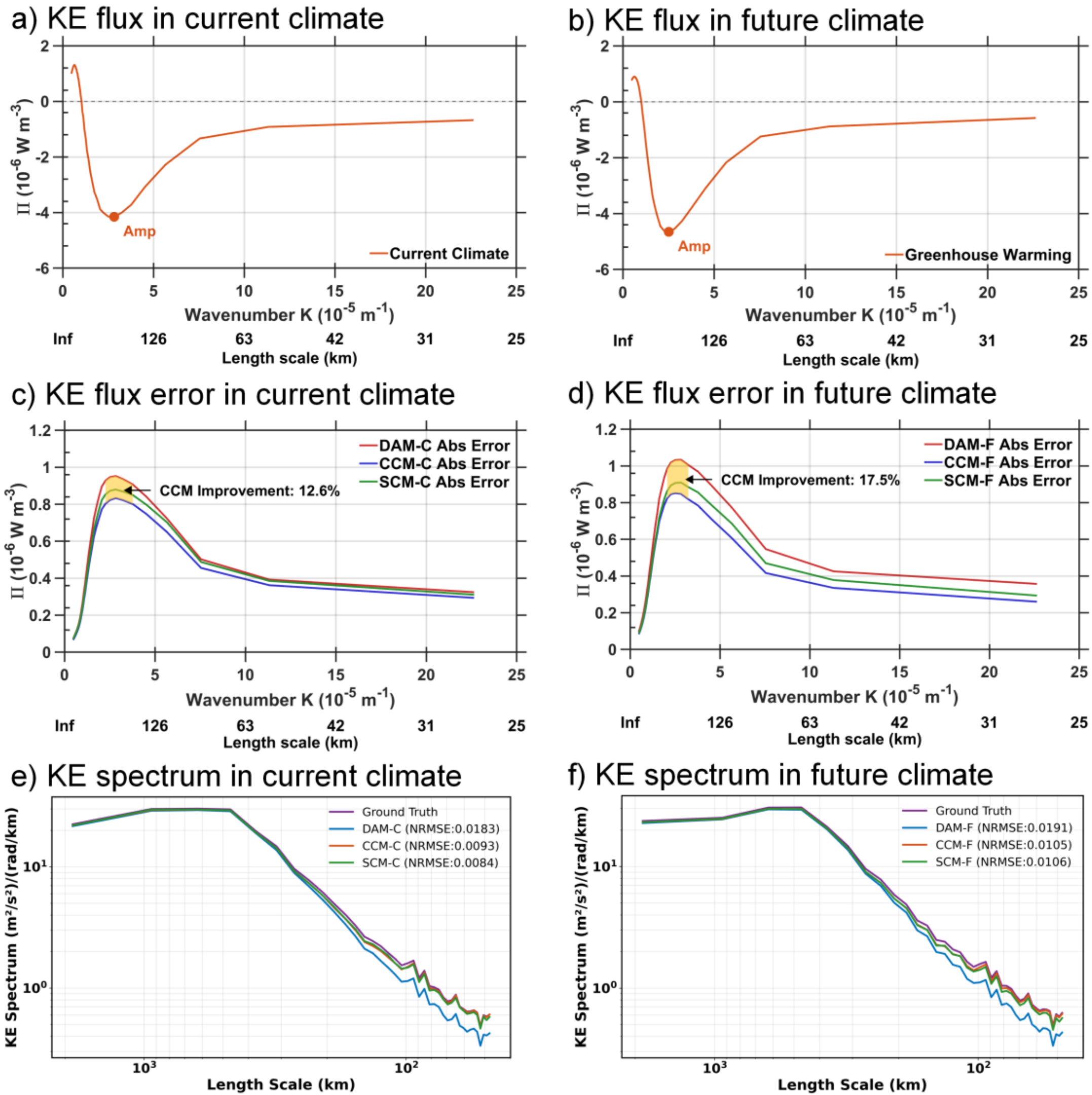

**Figure 8.** KE cascade spectra and model performance comparison. (a) Spectral KE flux Π(k) in the current climate, showing the characteristic amplitude minimum (Amp, red). (b) The same as (a), but for greenhouse warming scenario. (c) MAE between predicted and true spectral KE flux for the DAM-C (red), CCM-C (blue) and SCM-C (green) experiments. Yellow shading highlights the wavenumber range encompassing the five points with largest DAM-C absolute errors, where CCM shows 12.6% average improvement. (d) The same as (c), but for experiments in the future climate. (e) KE spectrum comparison in the current climate between the truth and the three experiments (DAM-C, CCM-C, SCM-C). (f) As in (e), but for the future-climate experiments (DAM-F, CCM-F, SCM-F).

A useful way to think about the KE spectrum and Π(k) is that the spectrum tells us how much energy resides at each scale, whereas the flux Π(k) quantifies the direction and strength of KE transfer across scales. Matching the spectral distribution does not automatically guarantee that the cross-scale transfer is correct, and the reverse is also possible. This motivates an explicit comparison between KE cascade-based and spectrum-based constraints. We therefore consider three configurations: the unconstrained DAM, the Cascade-Constrained model (CCM), and a Spectrum-Constrained model (SCM) that replaces the cascade loss used in CCM with a KE spectrum loss while keeping the remaining architecture and loss components unchanged. Both CCM and SCM reduce the absolute error of Π(k) relative to DAM in both climates (Figs. 8c,d), but CCM produces the largest improvement near the wavenumber band where DAM exhibits its maximum flux errors (yellow shading), with 12.6% improvement in the current climate and 17.5% under greenhouse warming. As to using the KE spectrum as constraints, though SCM also improves over DAM, the improvement is smaller (7.3% in the current climate and 12.1% under greenhouse warming).

The benefits of including physical constraints in ML models are also evident in the predicted KE spectrum (Fig. 8e,f). DAM underestimates KE at eddy scales in both climates, whereas CCM and SCM reproduce the reference spectrum more closely. Notably, CCM improves both KE spectrum and KE spectral flux, whereas SCM primarily targets the spectrum and yields a more limited improvement in the spectral flux (Figs. 8c–f). These results suggest that, compared to constraining the KE spectrum, constraining the cascade provides a more comprehensive improvement in energetic consistency than constraining the spectrum alone.

## 6 Conclusions

This study develops two ML models for SSH prediction and systematically evaluates their prediction skill and degree of climate adaptability. The novelty of DAM lies in integrating the dual attention mechanisms into ConvLSTM and developing a Trend-Magnitude Loss function, while CCM additionally introduces the KE cascade constraint through the ML training process. Both models maintain robust performance under altered oceanic conditions. Incorporating the KE cascade constraint can further improve the prediction skill of SSH and the relevant metrics (e.g., geostrophic EKE, skewness). Additional validations using AVISO observations and in the Gulf Stream region further confirm the robustness of the proposed models. A comparison with a KE-spectrum constraint further suggests that the cascade constraint improves both the KE flux and the KE spectrum, whereas the spectrum constraint primarily improves the KE spectrum.

This study provides a systematic and transferable framework for evaluating the climate adaptability of ML models. This framework could be applied across different ocean regions, datasets, climate scenarios and ML methods, which will be our future work. The KE cascade constraint introduced in this study, together with the dual attention mechanism, is architecture-agnostic and modular, allowing seamless integration into diverse ML architectures, including Transformers and Graph Neural Networks (Vaswani et al., 2017; Wu et al., 2020), opening a pathway for its application to emerging ML approaches in ocean modeling.

This work has two additional implications. One, the climate adaptability of ML models identified here suggests that ML models could possibly complement CMIP6 models in different climates. The appropriate merging of data-driven and physics-based models could eventually lead to an AI Model Intercomparison Project (AIMIP), which aims to predict and interpret climate change similar to CMIP. Two, we also find that the ML model can also reasonably represent the geostrophic KE cascade process. Therefore, similar to conventional numerical models, the ML model is not only useful for policy making purpose, but also potentially valuable for dynamical analysis.

## Conflict of Interest

The authors declare no conflicts of interest relevant to this study.

## Data Availability Statement

The daily SSH data from AWI-CM-1-1-MR used in this study are available from the World Data Center for Climate at DKRZ (https://doi.org/10.26050/WDCC/C6sCMAWAWM and https://doi.org/10.26050/WDCC/C6sSPAWAWM). The AVISO satellite altimetry data used in this study can be downloaded from https://data.marine.copernicus.eu/product/SEALEVEL-281GLO-PHY-L4-MY-008-047/description. The source code for the DAM and CCM models, along with training and testing scripts, is publicly available at https://github.com/Ezraocean/Climate-Adaptive-and-Cascade-Constrained-Machine-Learning-Prediction-for-Sea-Surface-Height.

## Acknowledgements

This study was funded by National Natural Science Foundation of China (Grant Nos. 42476007 and 42076007).

## References

Alexakis, A., & Biferale, L. (2018). Cascades and transitions in turbulent flows. *Physics Reports, 767*, 1-101.

Aluie, H., Hecht, M., & Vallis, G. K. (2018). Mapping the energy cascade in the North Atlantic Ocean: The coarse-graining approach. *Journal of Physical Oceanography, 48*(2), 225-244.

Bahdanau, D., Cho, K., & Bengio, Y. (2014). Neural machine translation by jointly learning to align and translate. *arXiv preprint arXiv:1409.0473*.

Barceló-Llull, B., Rosselló, P., Combes, V., Sánchez-Román, A., Pujol, M. I., & Pascual, A. (2025). Kuroshio Extension and Gulf Stream dominate the Eddy Kinetic Energy intensification observed in the global ocean. *Scientific Reports, 15*(1), 21754.

Beech, N., Rackow, T., Semmler, T., Danilov, S., Wang, Q., & Jung, T. (2022). Long-term evolution of ocean eddy activity in a warming world. *Nature climate change, 12*(10), 910-917.

Beucler, T., Gentine, P., Yuval, J., Gupta, A., Peng, L., Lin, J., Yu, S., Rasp, S., Ahmed, F., & O'Gorman, P. A. (2024). Climate-invariant machine learning. *Science Advances, 10*(6), eadj7250.

Bi, K., Xie, L., Zhang, H., Chen, X., Gu, X., & Tian, Q. (2023). Accurate medium-range global weather forecasting with 3D neural networks. *Nature, 619*(7970), 533-538.

Chelton, D. B., Schlax, M. G., & Samelson, R. M. (2011). Global observations of nonlinear mesoscale eddies. *Progress in oceanography, 91*(2), 167-216.

Chen, L., Zhong, X., Zhang, F., Cheng, Y., Xu, Y., Qi, Y., & Li, H. (2023). FuXi: A cascade machine learning forecasting system for 15-day global weather forecast. *npj climate and atmospheric science, 6*(1), 190.

Chen, Y., Wang, Y., Huang, G., & Tian, Q. (2024). Coupling physical factors for precipitation forecast in China with graph neural network. *Geophysical Research Letters, 51*(2), e2023GL106676.

Danilov, S., Kivman, G., & Schröter, J. (2004). A finite-element ocean model: principles and evaluation. *Ocean Modelling, 6*(2), 125-150.

Dramsch, J. S. (2020). 70 years of machine learning in geoscience in review. *Advances in geophysics, 61*, 1-55.

Eyring, V., Bony, S., Meehl, G. A., Senior, C. A., Stevens, B., Stouffer, R. J., & Taylor, K. E. (2016). Overview of the Coupled Model Intercomparison Project Phase 6 (CMIP6) experimental design and organization. *Geoscientific Model Development, 9*(5), 1937-1958.

Eyring, V., Collins, W. D., Gentine, P., Barnes, E. A., Barreiro, M., Beucler, T., Bocquet, M., Bretherton, C. S., Christensen, H. M., & Dagon, K. (2024). Pushing the frontiers in climate modelling and analysis with machine learning. *Nature climate change, 14*(9), 916-928.

Fox-Kemper, B. (2021). Ocean, cryosphere and sea level change. AGU fall meeting abstracts,

Geng, Q., Chen, R., Qiu, B., Jing, Z., Su, X., Huang, G., Liu, Q., & Chen, Y. (2025). Intensification of Oceanic Inverse Energy Cascade Under Global Warming. *arXiv preprint arXiv:2505.06327*.

Guillaumin, A. P., & Zanna, L. (2021). Stochastic-deep learning parameterization of ocean momentum forcing. *Journal of Advances in Modeling Earth Systems, 13*(9), e2021MS002534.

Gultekin, C., Subel, A., Zhang, C., Leibovich, M., Perezhogin, P., Adcroft, A., Fernandez-Granda, C., & Zanna, L. (2024). An Analysis of Deep Learning Parameterizations for Ocean Subgrid Eddy Forcing. *arXiv preprint arXiv:2411.06604*.

Ham, Y.-G., Kim, J.-H., & Luo, J.-J. (2019). Deep learning for multi-year ENSO forecasts. *Nature, 573*(7775), 568-572.

Hernanz, A., García-Valero, J. A., Domínguez, M., & Rodríguez-Camino, E. (2022). A critical view on the suitability of machine learning techniques to downscale climate change projections: Illustration for temperature with a toy experiment. *Atmospheric Science Letters, 23*(6), e1087.

Karpatne, A., Ebert-Uphoff, I., Ravela, S., Babaie, H. A., & Kumar, V. (2018). Machine learning for the geosciences: Challenges and opportunities. *IEEE Transactions on Knowledge and Data Engineering, 31*(8), 1544-1554.

Kashinath, K., Mustafa, M., Albert, A., Wu, J., Jiang, C., Esmaeilzadeh, S., Azizzadenesheli, K., Wang, R., Chattopadhyay, A., & Singh, A. (2021). Physics-informed machine learning: case studies for weather and climate modelling. *Philosophical Transactions of the Royal Society A, 379*(2194), 20200093.

Kochkov, D., Yuval, J., Langmore, I., Norgaard, P., Smith, J., Mooers, G., Klöwer, M., Lottes, J., Rasp, S., & Düben, P. (2024). Neural general circulation models for weather and climate. *Nature, 632*(8027), 1060-1066.

Kolmogorov, A. N. (1991). The local structure of turbulence in incompressible viscous fluid for very large Reynolds. *Numbers. In Dokl. Akad. Nauk SSSR, 30*, 301.

Li, G., Cheng, L., Zhu, J., Trenberth, K. E., Mann, M. E., & Abraham, J. P. (2020). Increasing ocean stratification over the past half-century. *Nature climate change, 10*(12), 1116-1123.

Lin, Z., Li, M., Zheng, Z., Cheng, Y., & Yuan, C. (2020). Self-attention convlstm for spatiotemporal prediction. Proceedings of the AAAI conference on artificial intelligence,

Ling, F., Luo, J.-J., Li, Y., Tang, T., Bai, L., Ouyang, W., & Yamagata, T. (2022). Multi-task machine learning improves multi-seasonal prediction of the Indian Ocean Dipole. *Nature Communications, 13*(1), 7681.

Martínez-Moreno, J., Hogg, A. M., England, M. H., Constantinou, N. C., Kiss, A. E., & Morrison, A. K. (2021). Global changes in oceanic mesoscale currents over the satellite altimetry record. *Nature climate change, 11*(5), 397-403.

Masson-Delmotte, V., Zhai, P., Pirani, A., Connors, S. L., Péan, C., Berger, S., Caud, N., Chen, Y., Goldfarb, L., & Gomis, M. (2021). Climate change 2021: the physical science basis. *Contribution of working group I to the sixth assessment report of the intergovernmental panel on climate change, 2*(1), 2391.

Meng, Y., Gao, F., Rigall, E., Dong, R., Dong, J., & Du, Q. (2023). Physical knowledge-enhanced deep neural network for sea surface temperature prediction. *IEEE Transactions on Geoscience and Remote Sensing, 61*, 1-13.

Meng, Y., Rigall, E., Chen, X., Gao, F., Dong, J., & Chen, S. (2021). Physics-guided generative adversarial networks for sea subsurface temperature prediction. *IEEE transactions on neural networks and learning systems, 34*(7), 3357-3370.

O'Gorman, P. A., & Dwyer, J. G. (2018). Using machine learning to parameterize moist convection: Potential for modeling of climate, climate change, and extreme events. *Journal of Advances in Modeling Earth Systems, 10*(10), 2548-2563.

O'Neill, B. C., Tebaldi, C., Van Vuuren, D. P., Eyring, V., Friedlingstein, P., Hurtt, G., Knutti, R., Kriegler, E., Lamarque, J.-F., & Lowe, J. (2016). The scenario model intercomparison project (ScenarioMIP) for CMIP6. *Geoscientific Model Development, 9*(9), 3461-3482.

Peng, Q., Xie, S.-P., Wang, D., Huang, R. X., Chen, G., Shu, Y., Shi, J.-R., & Liu, W. (2022). Surface warming–induced global acceleration of upper ocean currents. *Science Advances, 8*(16), eabj8394.

Perezhogin, P., Adcroft, A., & Zanna, L. (2025). Generalizable neural-network parameterization of mesoscale eddies in idealized and global ocean models. *Geophysical Research Letters, 52*(19), e2025GL117046.

Rasp, S., Pritchard, M. S., & Gentine, P. (2018). Deep learning to represent subgrid processes in climate models. *Proceedings of the national academy of sciences, 115*(39), 9684-9689.

Reichstein, M., Camps-Valls, G., Stevens, B., Jung, M., Denzler, J., Carvalhais, N., & Prabhat, F. (2019). Deep learning and process understanding for data-driven Earth system science. *Nature, 566*(7743), 195-204.

Rolnick, D., Donti, P. L., Kaack, L. H., Kochanski, K., Lacoste, A., Sankaran, K., Ross, A. S., Milojevic-Dupont, N., Jaques, N., & Waldman-Brown, A. (2022). Tackling climate change with machine learning. *ACM Computing Surveys (CSUR), 55*(2), 1-96.

Sallée, J.-B., Pellichero, V., Akhoudas, C., Pauthenet, E., Vignes, L., Schmidtko, S., Garabato, A. N., Sutherland, P., & Kuusela, M. (2021). Summertime increases in upper-ocean stratification and mixed-layer depth. *Nature, 591*(7851), 592-598.

Scott, R. B., & Wang, F. (2005). Direct evidence of an oceanic inverse kinetic energy cascade from satellite altimetry. *Journal of Physical Oceanography, 35*(9), 1650-1666.

Sein, D. V., Koldunov, N. V., Danilov, S., Wang, Q., Sidorenko, D., Fast, I., Rackow, T., Cabos, W., & Jung, T. (2017). Ocean modeling on a mesh with resolution following the local Rossby radius. *Journal of Advances in Modeling Earth Systems, 9*(7), 2601-2614.

Semmler, T., Danilov, S., Gierz, P., Goessling, H. F., Hegewald, J., Hinrichs, C., Koldunov, N., Khosravi, N., Mu, L., & Rackow, T. (2020). Simulations for CMIP6 with the AWI climate model AWI-CM-1-1. *Journal of Advances in Modeling Earth Systems, 12*(9), e2019MS002009.

Semmler, T., Danilov, S., Rackow, T., Sidorenko, D., Barbi, D., Hegewald, J., Pradhan, H. K., Sein, D., Wang, Q., & Jung, T. (2019). IPCC DDC: AWI AWI-CM1. 1MR model output prepared for CMIP6 ScenarioMIP ssp585.

Sérazin, G., Penduff, T., Barnier, B., Molines, J.-M., Arbic, B. K., Müller, M., & Terray, L. (2018). Inverse cascades of kinetic energy as a source of intrinsic variability: A global OGCM study. *Journal of Physical Oceanography, 48*(6), 1385-1408.

Shannon, C. E. (1948). A mathematical theory of communication. *The Bell system technical journal, 27*(3), 379-423.

Shi, X., Chen, Z., Wang, H., Yeung, D.-Y., Wong, W.-K., & Woo, W.-c. (2015). Convolutional LSTM network: A machine learning approach for precipitation nowcasting. *Advances in neural information processing systems, 28*.

Song, T., Han, R., Meng, F., Wang, J., Wei, W., & Peng, S. (2022). A significant wave height prediction method based on deep learning combining the correlation between wind and wind waves. *Frontiers in Marine Science, 9*, 983007.

Stammer, D. (1997). Global characteristics of ocean variability estimated from regional TOPEX/POSEIDON altimeter measurements. *Journal of Physical Oceanography, 27*(8), 1743-1769.

Storer, B. A., Buzzicotti, M., Khatri, H., Griffies, S. M., & Aluie, H. (2023). Global cascade of kinetic energy in the ocean and the atmospheric imprint. *Science Advances, 9*(51), eadi7420.

Su, Z., Wang, J., Klein, P., Thompson, A. F., & Menemenlis, D. (2018). Ocean submesoscales as a key component of the global heat budget. *Nature Communications, 9*(1), 775.

Sura, P., & Gille, S. T. (2010). Stochastic dynamics of sea surface height variability. *Journal of Physical Oceanography, 40*(7), 1582-1596.

Vallis, G. K. (2017). *Atmospheric and oceanic fluid dynamics*. Cambridge University Press.

Vaswani, A., Shazeer, N., Parmar, N., Uszkoreit, J., Jones, L., Gomez, A. N., Kaiser, Ł., & Polosukhin, I. (2017). Attention is all you need. *Advances in neural information processing systems, 30*.

Von Storch, H., & Zwiers, F. W. (2002). *Statistical analysis in climate research*. Cambridge university press.

Wang, Q., Danilov, S., Sidorenko, D., Timmermann, R., Wekerle, C., Wang, X., Jung, T., & Schröter, J. (2014). The Finite Element Sea Ice-Ocean Model (FESOM) v. 1.4: formulation of an ocean general circulation model. *Geoscientific Model Development, 7*(2), 663-693.

Wang, S., Jing, Z., Wu, L., Sun, S., Chen, Z., Ma, X., & Gan, B. (2024). A more quiescent deep ocean under global warming. *Nature climate change, 14*(9), 961-967.

Wang, S., Liu, Z., & Pang, C. (2015). Geographical distribution and anisotropy of the inverse kinetic energy cascade, and its role in the eddy equilibrium processes. *Journal of Geophysical Research: Oceans, 120*(7), 4891-4906.

Wang, X., Wang, R., Hu, N., Wang, P., Huo, P., Wang, G., Wang, H., Wang, S., Zhu, J., & Xu, J. (2024). Xihe: A data-driven model for global ocean eddy-resolving forecasting. *arXiv preprint arXiv:2402.02995*.

Wu, Z., Pan, S., Chen, F., Long, G., Zhang, C., & Yu, P. S. (2020). A comprehensive survey on graph neural networks. *IEEE transactions on neural networks and learning systems, 32*(1), 4-24.

Yuval, J., & O'Gorman, P. A. (2020). Stable machine-learning parameterization of subgrid processes for climate modeling at a range of resolutions. *Nature Communications, 11*(1), 3295.

Zhang, G., Chen, R., Li, X., Li, L., Wei, H., & Guan, W. (2023). Temporal variability of global surface eddy diffusivities: Estimates and machine learning prediction. *Journal of Physical Oceanography, 53*(7), 1711-1730.

Zheng, L., Lu, W., & Zhou, Q. (2023). Weather image-based short-term dense wind speed forecast with a ConvLSTM-LSTM deep learning model. *Building and Environment, 239*, 110446.

Zhou, G., & Cheng, X. (2022). Impacts of oceanic fronts and eddies in the Kuroshio-Oyashio Extension region on the atmospheric general circulation and storm track. *Advances in Atmospheric Sciences, 39*(1), 22-54.

Zhu, Y., Zhang, R.-H., Moum, J. N., Wang, F., Li, X., & Li, D. (2022). Physics-informed deep-learning parameterization of ocean vertical mixing improves climate simulations. *National Science Review, 9*(8), nwac044.